\documentclass[a4paper,11pt]{article}

\usepackage{jheppub} 
\usepackage[normalem]{ulem}  
\usepackage{amsmath,amssymb}

\newcommand{\bsk}{{\boldsymbol{k}}}
\newcommand{\bsx}{{\boldsymbol{x}}}

\newcommand{\nn}{\nonumber}

\title{
Wavefunctions in dS/CFT revisited:
\\
principal series and double-trace deformations
}

\author[a]{Hiroshi Isono,}
\author[b]{Hoiki Madison Liu}
\author[c]{and Toshifumi Noumi}

\affiliation[a]{Department of Physics, Faculty of Science, Chulalongkorn University, Bangkok 10330, Thailand}
\affiliation[b]{Institute of Physics, University of Tokyo, Tokyo 153-8092, Japan}
\affiliation[c]{Department of Physics, Kobe University, Kobe 657-8501, Japan}

\emailAdd{hiroshi.isono81@gmail.com}
\emailAdd{h.liu@hep1.c.u-tokyo.ac.jp}
\emailAdd{tnoumi@phys.sci.kobe-u.ac.jp}

\preprint{UT-Komaba/20-4, KOBE-COSMO-20-11}

\abstract{
We study wavefunctions of heavy scalars on de Sitter spacetime and their implications to dS/CFT correspondence. In contrast to light fields in the complementary series, heavy fields in the principal series oscillate outside the cosmological horizon. As a consequence, the quadratic term in the wavefunction does not follow a simple scaling and so it is hard to identify it with a conformal two-point function. In this paper, we demonstrate that it should be interpreted as a two-point function on a cyclic RG flow which is obtained by double-trace deformations of the dual CFT. This is analogous to the situation in nonrelativistic AdS/CFT with a bulk scalar whose mass squared is below the Breitenlohner-Freedman (BF) bound. We also provide a new dS/CFT dictionary relating de Sitter two-point functions and conformal two-point functions in the would-be dual CFT.
}

\begin{document} 
\setcounter{tocdepth}{2}
\maketitle
\flushbottom

\section{Introduction and summary}

The dS/CFT correspondence postulates that wavefunctions on $(d+1)$-dimensional de Sitter space are equivalent at the future infinity to generating functions of the dual CFT in $d$ dimensions~\cite{Maldacena:2002vr}. In momentum space (along the CFT direction), it claims that\footnote{
Primed correlators are defined by $\langle\ldots\rangle=(2\pi)^d\delta^{d}(\sum_i\bsk_i)\langle\ldots\rangle'$.}
\begin{align}
\Psi[\bar{\phi}]=\exp\left[
\frac{1}{2}\int \frac{d^d\bsk}{(2\pi)^d} 
\langle O_{\bsk}O_{-\bsk}\rangle'\,\bar{\phi}_{\bsk}\bar{\phi}_{-\bsk}+\mathcal{O}(\bar{\phi}^3)
\right]\,,
\end{align}
where $\Psi[\bar{\phi}]$ is the wavefunction of a bulk scalar field $\phi$, $\bar{\phi}$ is its boundary value, and $O_{\bsk}$ is the dual CFT operator. We focus on bulk free theories, so that we neglect the $\mathcal{O}(\bar{\phi}^3)$ terms in the wavefunction. This provides a famous dS/CFT dictionary:
\begin{align}
\label{dS/CFT_2pt}
\langle\phi_{\bsk}(\tau_*)\phi_{-\bsk}(\tau_*)\rangle'=-\frac{1}{2\,{\rm Re}\,\langle O_{\bsk}O_{-\bsk}\rangle'}\,,
\end{align}
where $\tau_*$ is a late time the wavefunction is evaluated. For example, two-point functions of light scalars with the mass $0\leq m<\frac{d}{2}H$ in the superhorizon regime $-k\tau_*\ll1$ read
\begin{align}
\label{2pt_light}
\langle\phi_{\bsk}(\tau_*)\phi_{-\bsk}(\tau_*)\rangle'\propto (-\tau_*)^d(-k\tau_*)^{-2\nu}
\quad
{\rm with}
\quad
\nu=\sqrt{\frac{d^2}{4}-\frac{m^2}{H^2}}\,,
\end{align}
where $H$ is the Hubble constant, i.e., inverse of the de Sitter radius. This can naturally be identified with the inverse of two-point functions of the dual CFT operator with the scaling dimension $\Delta=\frac{d}{2}+\nu$.

\medskip
The purpose of this paper is to revisit the dictionary~\eqref{dS/CFT_2pt} for heavy scalars in the principal series. In contrast to light fields, heavy fields oscillate outside the cosmological horizon and so do their two-point functions:
\begin{align}
\label{2pt_heavy}
\langle\phi_{\bsk}(\tau_*)\phi_{-\bsk}(\tau_*)\rangle'\propto(-\tau_*)^d \left|1+e^{-\pi\mu}e^{i\alpha(\mu)}(-k\tau_*)^{2i\mu}\right|^2
\quad
{\rm with}
\quad
\mu=\sqrt{\frac{m^2}{H^2}-\frac{d^2}{4}}\,,
\end{align}
where $\alpha(\mu)$ is a mass-dependent phase factor. At least naively, it is hard to identify them with inverse of (a real part  of) conformal two-point functions. To our knowledge, most of the dS/CFT literature~\cite{Spradlin:2001nb,Larsen:2002et,Maldacena:2002vr,Larsen:2003pf,vanderSchaar:2003sz,Seery:2006tq,McFadden:2009fg,McFadden:2010na,McFadden:2010vh,Harlow:2011ke,McFadden:2011kk,Anninos:2011ui,Bzowski:2011ab,Das:2012dt,Anninos:2012ft,Smolkin:2012er,Schalm:2012pi,Bzowski:2012ih,Mata:2012bx,Garriga:2013rpa,Anninos:2013rza,Das:2013qea,Banerjee:2013mca,McFadden:2013ria,Pimentel:2013gza,Kol:2013msa,Ghosh:2014kba,Xiao:2014uea,Garriga:2014ema,Anninos:2014lwa,Kundu:2014gxa,Garriga:2014fda,Fei:2015kta,Kundu:2015xta,Shukla:2016bnu,Isono:2016yyj,Hertog:2017ymy,Neiman:2017zdr,Anninos:2017eib,Hui:2018cag,Hertog:2019uhy,Sengor:2019mbz,Yokoyama:2020tgs,David:2020ptn,Heckelbacher:2020nue} including higher spin holography and holographic inflation have focused on light fields, and so this issue has not been well addressed. However, such heavy fields are inevitable once we consider UV completion of the bulk theory. In this paper, we demonstrate that they should be interpreted as two-point functions on a cyclic RG flow generated by double-trace deformations, similarly to the AdS tachyon with the mass squared below the Breitenlohner-Freedman (BF) bound~\cite{Moroz:2009kv}.

\begin{table}[t]
\begin{center}
\renewcommand\arraystretch{1.5}
\newcommand{\bhline}[1]{\noalign{\hrule height #1}}
\begin{tabular}{|c||c|c|c|}
\hline
AdS/CFT & $m^2/H_{\rm AdS}^2>-\frac{d^2}{4}+1$ & $-\frac{d^2}{4}<m^2/H_{\rm AdS}^2<-\frac{d^2}{4}+1$ & $m^2/H_{\rm AdS}^2<-\frac{d^2}{4}$ \\
\bhline{1.5pt}
$\Delta_+$ & $\Delta_+>\frac{d}{2}+1$&$\frac{d}{2}<\Delta_+<\frac{d}{2}+1$&$\Delta_+=\frac{d}{2}+i\mu_{\rm AdS}$\\
\hline
$\Delta_-$ & $\Delta_-<\frac{d}{2}-1$&$\frac{d}{2}-1<\Delta_-<\frac{d}{2}$&$\Delta_-=\frac{d}{2}-i\mu_{\rm AdS}$\\
\hline
\end{tabular}
\vspace{1mm}
\caption{
AdS scalars have two modes with the asymptotic behavior $\phi\sim z^{\Delta_\pm}$ near the AdS boundary, where $\Delta_\pm=\frac{d}{2}\pm\sqrt{\frac{m^2}{H_{\rm AdS}^2}+\frac{d^2}{4}}$. Canonical quantization in AdS shows that the mode $\phi\sim z^\Delta$ has a finite energy only when $\Delta>\frac{d}{2}-1$, which coincides with the unitarity bound in the dual CFT. When the mass squared is below the BF bound $m^2<-\frac{d^2}{4}$, there appears tachyonic instability. Also, in this regime, the dual CFT operator has a complex conformal dimension $\Delta_\pm=\frac{d}{2}\pm i\mu_{\rm AdS}$ with $\mu_{\rm AdS}=\sqrt{-\frac{m^2}{H_{\rm AdS}^2}-\frac{d^2}{4}}$, which is prohibited by unitarity in the dual CFT.}
\label{table:AdS/CFT}
\vspace{-3mm}
\end{center}
\end{table}

\medskip
Before proceeding to details of the discussion, it is convenient to summarize our results in the connection with the story in AdS/CFT (see also Table~\ref{table:AdS/CFT}). For this purpose, let us recall boundary conditions for quantization of scalar fields on AdS~\cite{Breitenlohner:1982jf,Balasubramanian:1998sn,Klebanov:1999tb}. First, when the mass squared $m^2$ is in the range $m^2>(-\frac{d^2}{4}+1)H_{\rm AdS}^2$, the unique admissible boundary condition is the Dirichlet boundary condition that kills the mode with the asymptotic behavior,
\begin{align}
\phi\sim z^{\Delta_-} \,,\quad \Delta_\pm=\frac{d}{2}\pm\sqrt{\frac{m^2}{H_{\rm AdS}^2}+\frac{d^2}{4}}\,,
\end{align}
where $H_{\rm AdS}$ is inverse of the AdS radius and $z$ is the scale direction of the Poincare coordinate ($z=0$ is the AdS boundary). For $-\frac{d^2}{4}H_{\rm AdS}^2<m^2<(-\frac{d^2}{4}+1)H_{\rm AdS}^2$, on the other hand, there are two possible boundary conditions essentially because the two modes $\phi\sim z^{\Delta_\pm}$ carry a finite energy. In particular, one may employ a mixed boundary condition that kills the mode $\phi\sim z^{\Delta_+}$. It is known that the CFT dual to the mixed boundary condition is related to the CFT dual to the Dirichlet boundary condition through an RG flow induced by double-trace deformations~\cite{Witten:2001ua,Berkooz:2002ug,Gubser:2002zh,Gubser:2002vv,Hartman:2006dy}\footnote{The RG induced by double-trace deformations are also discussed in~\cite{Heemskerk:2010hk,Faulkner:2010jy,Elander:2011vh} in the context of Wilsonian holographic renormalization group.}. Note that, as is known as the BF bound, tachyonic instability appears in the mass range $m^2<-\frac{d^2}{4}H_{\rm AdS}^2$. Correspondingly, the dual CFT operator has a complex conformal dimension, which violates unitarity.

\medskip
Now let us get back to the dS story. First, let us recall that in the computation of dS correlators, we do not impose any boundary condition at the future infinity, but rather we impose boundary conditions deep inside the horizon alone. This reflects the closed time path nature of the in-in formalism (the Schwinger-Keldysh formalism). On the other hand, the dS/CFT dictionary equates dS wavefunctions at late time with generating functions of the would-be dual CFTs. While we normally employ the Dirichlet boundary condition for the computation of wavefunctions, there is no a priori reason for this: a different choice of boundary conditions simply means a different representation of wavefunctions, so that in contrast to the AdS case, there is no physical criterion to select boundary conditions. Indeed, in the context of higher spin dS/CFT~\cite{Anninos:2011ui,Anninos:2013rza}, one may consider two different boundary conditions dual to the free $Sp(N)$ model and the critical $Sp(N)$ model, which are interpolated via an RG flow induced by double-trace deformations~\cite{Anninos:2013rza}. See also Ref.~\cite{Das:2013qea} for double-trace deformations in dS/CFT with bulk scalars in the complementary series.

\begin{table}[t]
\begin{center}
\renewcommand\arraystretch{1.5}
\newcommand{\bhline}[1]{\noalign{\hrule height #1}}
\begin{tabular}{|c||c|c|c|}
\hline
dS/CFT & $m^2/H^2<0$ & $\frac{d^2}{4}>m^2/H^2>0$ & $m^2/H^2>\frac{d^2}{4}$ \\
\bhline{1.5pt}
$\Delta_+$ & $\Delta_+>d$&$\frac{d}{2}<\Delta_+<d$&$\Delta_+=\frac{d}{2}+i\mu$\\
\hline
$\Delta_-$ & $\Delta_-<0$&$0<\Delta_-<\frac{d}{2}$&$\Delta_-=\frac{d}{2}-i\mu$\\
\hline
\end{tabular}
\vspace{1mm}
\caption{
Scalar fields on dS have two modes with the asymptotic late time behavior $\phi\sim (-\tau)^{\Delta_\pm}$, where $\Delta_\pm=\frac{d}{2}\pm\sqrt{\frac{d^2}{4}-\frac{m^2}{H^2}}$ for $m^2<\frac{d^2}{4}H^2$ and $\Delta_\pm=\frac{d}{2}\pm i\mu$ for $m^2>\frac{d^2}{4}H^2$. Unitary representations on de Sitter space are classified into the principal series $m^2>\frac{d^2}{4}H^2$ and the complementary series $0<m^2<\frac{d^2}{4}H^2$. In contrast to the AdS/CFT case, the time direction in the bulk is not along the CFT direction, so that unitarity in the bulk does not imply that in the would-be dual CFT. In particular, for the principal series, the dual CFT operator has a complex conformal dimension, similarly to AdS scalars with the mass squared below the BF bound.}
\label{table:dS/CFT}
\vspace{-3mm}
\end{center}
\end{table}

\medskip
In this paper, we study a similar problem for heavy scalars in the principal series. First, we point out that for heavy fields, a mixed boundary condition gives a natural identification of dS wavefunctions with generating functions of the would-be dual CFTs. In particular, QFTs dual to wavefunctions with the Dirichlet boundary condition are their double-trace deformations, similarly to the AdS story mentioned earlier.
Another property characteristic to heavy fields on dS is that the RG flow generated by double-trace deformations is cyclic, simply because the dual CFT operator has a complex conformal dimension (see Table~\ref{table:dS/CFT}). This situation is analogous to AdS tachyons which violate the BF bound: Recall that dS and AdS are related via the analytic continuation $H^2\to -H_{\rm AdS}^2$ together with double Wick rotations. Therefore, AdS scalars corresponding to dS scalars in the principal series $m^2/H^2>\frac{d^2}{4}$ have the mass squared  $m^2/H_{\rm AdS}^2<-\frac{d^2}{4}$, which violates the BF bound. Such fields are prohibited in the standard AdS/CFT, but Ref.~\cite{Moroz:2009kv} argued that they are acceptable in nonrelativistic AdS/CFT and studied their relation to RG limit cycles. In dS/CFT, essentially the same situation appears even without considering nonrelativistic setups, simply by taking heavy fields in the bulk which are inevitable when we consider UV completion of the bulk theory.

\medskip
Besides, based on these observations on dS wavefunctions, we provide a dictionary between dS two-point functions and CFT two-point functions of the form,
\begin{align}
\label{dS/CFT_mixed}
\langle \phi_\bsk (\tau_* )\phi_{-\bsk} (\tau_* )\rangle '&=
-(-\tau_*)^{2d}\frac{\langle O_\bsk O_{-\bsk} \rangle '_{M+}\langle O_\bsk O_{-\bsk} \rangle '_{M-}}{\langle O_\bsk O_{-\bsk} \rangle '_{M+}+\langle O_\bsk O_{-\bsk} \rangle '_{M-}}\,,
\end{align}
where $\langle O_\bsk O_{-\bsk} \rangle '_{M\pm}$ are two-point functions in QFTs dual to wavefunctions with mixed boundary conditions. The subscripts $\pm$ are associated to the time ordered and anti-time ordered integration contours in the in-in formalism (see Sec.~\ref{Sec:wavefunctions} for details).  For light scalars in the complementary series, $\langle O_\bsk O_{-\bsk} \rangle '_{M\pm}$ coincide with conformal two-point functions by themselves. On the other hand, for heavy scalars in the principal series, they coincide with conformal two-point functions up to analytic terms which can be subtracted by local counterterms in a similar manner to holographic renormalization in the AdS/CFT context~\cite{Bianchi:2001kw,Skenderis:2002wp}. This provides a new dS/CFT dictionary applicable when mixed boundary conditions are employed.

\medskip
The rest of the paper is organized as follows: In Sec.~\ref{Sec:wavefunctions}, we demonstrate that dS wavefunctions with mixed boundary conditions are naturally identified with generating functions of the would-be dual CFTs. Then, we show that wavefunctions with the Dirichlet boundary conditions are dual to their double-trace deformations. In Sec.~\ref{Sec:correlators}, we provide a dS/CFT dictionary~\eqref{dS/CFT_mixed} applicable when the mixed boundary conditions are employed.  We conclude in Sec.~\ref{Sec:discussion} with discussion of our results. Some technical details are given in Appendix.

\section{Wavefunctions on de Sitter space}
\label{Sec:wavefunctions}

In this section we revisit holographic interpretation of dS wavefunctions, mainly focusing on heavy scalars in the principal series. First, we show that for heavy fields, wavefunctions evaluated with mixed boundary conditions are naturally identified with a generating function of the would-be dual CFT. Then, we demonstrate that wavefunctions evaluated with the Dirichlet boundary conditions are identified with generating functions of QFTs on a cyclic RG flow obtained by double-trace deformations. For comparison, we also discuss mixed boundary conditions for light fields at the end of the section, following Ref.~\cite{Das:2013qea}.

\subsection{Dirichlet boundary conditions}

We begin by the Dirichlet boundary problem on de Sitter space, which is normally used to calculate dS correlators directly from the wavefunction.
Let us consider a free scalar $\phi$ on $(d+1)$-dimensional de Sitter space with the unit radius:
\begin{align}
\nonumber
S[\phi]&=
-\frac{1}{2}\int d^{d+1}x\sqrt{-g}\Big[
(\partial_\mu\phi)^2+m^2\phi^2
\Big]
\\
\label{dS_action_D_pre}
&
=\frac{1}{2}\int\frac{d\tau}{(-\tau)^{d+1}}\int \frac{d^d\bsk}{(2\pi)^d}\Big[
(\theta_\tau\phi_{-\bsk})(\theta_\tau\phi_{\bsk})
-(m^2+k^2\tau^2)\phi_{-\bsk}\phi_{\bsk}
\Big]
\,.
\end{align}
Here and in what follows we use the planar coordinates,
\begin{align}
ds^2=\frac{-d\tau^2+d\bsx^2}{\tau^2}\,,
\end{align}
and work in momentum space along the CFT direction $\bsx$. Also, $\theta_x=x\frac{\partial}{\partial x}$ is the Euler operator, counting the power in $x$.

\begin{figure}[t]
\begin{center}
\includegraphics[width=75mm, bb=0 0 842 595]{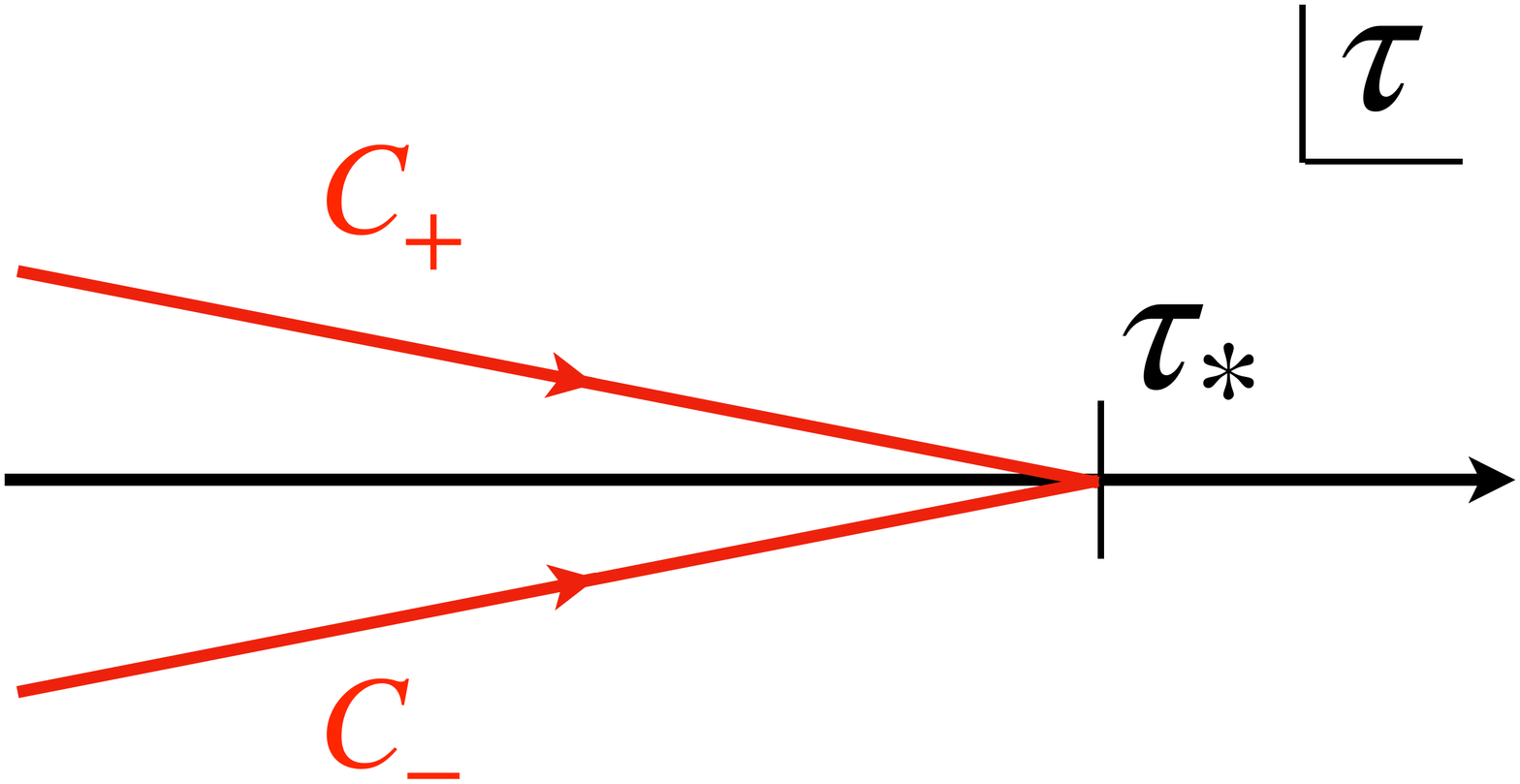}
\end{center}
\vspace{-10mm}
\caption{Time integration contours $\mathcal{C}_\pm$: The path integral measure in the in-in formalism is given by $\exp (i\int_{\mathcal{C}_+-\mathcal{C}_-}\!\!d\tau \int d^{d}\bsx \sqrt{-g}\,\mathcal{L})=\exp (iS_+-iS_-)$, where we defined $S_\pm=\int_{\mathcal{C}_\pm}d\tau \int d^d\bsx \sqrt{-g}\,\mathcal{L}$. In particular, $\mathcal{C}_+-\mathcal{C}_-$ forms a closed time path. We call $\mathcal{C}_+$ and $\mathcal{C}_-$ the time ordered path and the anti-time ordered path, respectively.}
\label{Fig:CTP}
\end{figure}

\medskip
To follow the dynamics on de Sitter space, we use the in-in formalism (the Schwinger-Keldysh formalism), where the time integral is defined along a closed path. As depicted in Fig.~\ref{Fig:CTP}, the time ordered contour $\mathcal{C}_+$ and the anti-time ordered contour $\mathcal{C}_-$ extend over $(-(1- i\epsilon)\infty,\tau_*]$ and $(-(1+ i\epsilon)\infty,\tau_*]$, respectively. Here $\tau_*$ is a late time near the future boundary. We denote the action defined with each integration contour by
\begin{align}
\label{dS_action_D}
S^{\pm}[\phi]&=
\frac{1}{2}\int_{\mathcal{C}_{\pm}}\frac{d\tau}{(-\tau)^{d+1}}\int \frac{d^d\bsk}{(2\pi)^d}\Big[
(\theta_\tau\phi_{-\bsk})(\theta_\tau\phi_{\bsk})
-(m^2+k^2\tau^2)\phi_{-\bsk}\phi_{\bsk}
\Big]
\,.
\end{align}
In this language, the de Sitter wavefunction is defined by
\begin{align}
\Psi^{\pm}_D[\bar{\phi}]=\int_{\bar{\phi}, \mathrm{BD}_{\pm}}[d \phi] e^{\pm i S^{\pm}[\phi]}\,,
\end{align}
where the path integral for $\Psi^{\pm}_D[\bar{\phi}]$ is performed with the Dirichlet boundary conditions at $\tau=\tau_*$ and the Bunch-Davies vacuum conditions at the past infinity:
\begin{align}
\label{bc_dS}
\phi_{\bsk}(\tau_*)=\bar{\phi}_{\bsk}\,,
\quad
\lim_{\tau\to -(1\mp i\epsilon)\infty}\phi_{\bsk}(\tau)=0\,.
\end{align}
Here $\bar{\phi}$ is a boundary scalar field. Throughout the paper, we use the semiclassical approximation, under which the wavefunction is evaluated with the on-shell action as
\begin{align}
\label{wavefunction_def}
\Psi^\pm_D[\bar{\phi}]=e^{ \pm iS^\pm [\phi^\pm]}
\quad
{\rm with}
\quad
S^\pm[\phi^\pm ]=-\frac{1}{2}\int \frac{d^d\bsk}{(2\pi)^d}\left[(-\tau)^{-d}\phi^\pm_{ -\bsk}\theta_\tau\phi^\pm _{\bsk}
\right]_{\tau=-(1\mp i\epsilon)\infty}^{\tau_*}
\,,
\end{align}
where $\phi^\pm $ is a solution of the equation of motion,
\begin{align}
\label{eom_dS}
\left[
\theta_\tau(\theta_\tau-d)
+(m^2+k^2\tau^2)
\right]\phi^\pm_{\bsk}=0\,,
\end{align}
which satisfies the boundary conditions \eqref{bc_dS}. Note that the on-shell Lagrangian vanishes up to total derivatives and so the on-shell action is localized at the boundaries as usual.

\medskip
The solution $\phi^\pm$ is given explicitly by
\begin{align}
\label{dS_solution}
\phi^\pm _{\bsk}(\tau)=\mathcal{B}_D^\pm (k;\tau)\bar{\phi}_{\bsk}\,,
\end{align}
where $\mathcal{B}^\pm_D$ are the bulk-to-boundary propagators for each time path. For heavy fields in the principal series $m>\frac{d}{2}$ (see Sec.~\ref{subsec:light} for light fields), we have
\begin{align}
\mathcal{B}_D^+ (k;\tau)=\frac{(-\tau)^{\frac{d}{2}}H^{(2)}_{-i\mu}(-k\tau)}{(-\tau_*)^{\frac{d}{2}}H^{(2)}_{-i\mu}(-k\tau_*)}\,, \quad
\mathcal{B}_D^- (k;\tau)=\frac{(-\tau)^{\frac{d}{2}}H^{(1)}_{i\mu}(-k\tau)}{(-\tau_*)^{\frac{d}{2}}H^{(1)}_{i\mu}(-k\tau_*)}\,,
\end{align}
where we introduced $\mu=\sqrt{{m^2-\frac{d^2}{4}}}$.
In the superhorizon regime $-k\tau_*\leq -k\tau\ll1$, the bulk-to-boundary propagator behaves as
\begin{align}
\label{dS_heavy_B}
\mathcal{B}_D^\pm (k;\tau)&\simeq(\tau/\tau_*)^{\frac{d}{2}\mp i\mu}\frac{1+e^{- \pi\mu}e^{ \pm i\alpha(\mu)}(-k\tau)^{ \pm 2i\mu}}{1+e^{- \pi\mu}e^{\pm i\alpha(\mu)}(-k\tau_*)^{ \pm 2i\mu}}
\quad
{\rm with}
\quad
e^{i\alpha(\mu)}=\frac{\Gamma(-i\mu)}{2^{2i\mu}\Gamma(i\mu)}
\,,
\end{align}
where we introduced the mass-dependent phase factor $\alpha(\mu)$. Note that $\mathcal{B}_D^-$ is the complex conjugate of $\mathcal{B}_D^+$. Also, for $\mathcal{B}_D^+$, the first and second terms in the numerator describe the positive and negative frequency modes at late time, respectively, and vice versa for $\mathcal{B}_D^-$. The factor $e^{-\pi\mu}e^{\pm i\alpha(\mu)}$ is nothing but the one appearing in the thermal Bogoliubov coefficients.

\medskip
Now we evaluate the wavefunction. Substituting the solution~\eqref{dS_solution} into Eq.~\eqref{wavefunction_def} gives
\begin{align}
\Psi^\pm_D [\bar{\phi}]&=\exp
\left[
\mp\frac{i}{2}\int \frac{d^d\bsk}{(2\pi)^d}
(-\tau_*)^{-d}\,
\left[\theta_\tau\mathcal{B}_D^\pm(k;\tau)\right]_{\tau=\tau_*}
\bar{\phi}_{\bsk}\bar{\phi}_{-\bsk}
\right]
\nonumber
\\
\label{wavefunction_dS_D}
&\simeq\exp
\left[
\mp \frac{1}{2}\int \frac{d^d\bsk}{(2\pi)^d}
(-\tau_*)^{-d}\,
\left(
\frac{d}{2}i +\mu \cdot\frac{1-e^{\mp \pi\mu}e^{ i\alpha(\mu)}(-k\tau_*)^{ 2i\mu}}{1+e^{\mp\pi\mu}e^{ i\alpha(\mu)}(-k\tau_*)^{ 2i\mu}}
\right)
\bar{\phi}_{\bsk}\bar{\phi}_{-\bsk}
\right]\,.
\end{align}
The holographic dictionary then states that
\begin{align}
\label{OO_Dirichlet}
\langle O_{\bsk}O_{-\bsk}\rangle'_{D\pm} =\mp (-\tau_*)^{-d}\left[\frac{d}{2}i +\mu \cdot\frac{1-e^{\mp\pi\mu}e^{ i\alpha(\mu)}(-k\tau_*)^{ 2i\mu}}{1+e^{\mp\pi\mu}e^{ i\alpha(\mu)}(-k\tau_*)^{ 2i\mu}}\right]\,,
\end{align}
where $\langle O_{\bsk}O_{-\bsk}\rangle'_{D\pm}$ are the two-point functions of the dual QFT operators $O$ and the prime implies that the delta function for momentum conservation is dropped. Also the subscripts $D$ and $\pm$ specify the boundary condition and the time path.
Essentially because of the oscillatory behavior of the heavy field outside the horizon, it is hard to identify Eq.~\eqref{OO_Dirichlet} with a conformal two-point function which respects the dilatation Ward-Takahashi identity\footnote{
Note that there is a freedom to add a local boundary action (or equivalently, total derivative terms in Lagrangian), which corresponds to renormalization in the dual QFT~\cite{Bianchi:2001kw,Skenderis:2002wp}. At the level of our interest, this is equivalent with adding a pure imaginary constant to Eq.~\eqref{OO_Dirichlet}, but obviously it does not help. Also, this freedom does not change the wavefunction squared $|\Psi|^2$ and so the result of canonical quantization is of course reproduced for whatever choice of holographic renormalization scheme:
\begin{align}
\langle \phi_{\bsk}(\tau_*)\phi_{-\bsk}(\tau_*)\rangle'&=
\frac{(-\tau_*)^{d}}{2\mu}\cdot\frac{|1+e^{-\pi\mu}e^{i\alpha(\mu)}(-k\tau_*)^{2i\mu}|^2}{1-e^{-2\pi\mu}}\,.\label{canonical2pt}
\end{align}
}. In the rest of the section, we shall demonstrate that it should be identified with a two-point functions on a cyclic RG flow induced by a double-trace deformation.

\subsection{Mixed boundary conditions}

Next, we show that wavefunctions with mixed boundary conditions are naturally identified with CFT generating functions. As we have discussed, in the Dirichlet boundary problem, the two modes $\phi\sim(-\tau)^{\frac{d}{2}\pm i\mu}$ mix with each other at late time, which was the main obstruction to identifying dS wavefunctions with CFT generating functions. Here we employ the following mixed boundary conditions instead:
\begin{align}
\label{mixed_bc}
\left(\theta_\tau-\tfrac{d}{2}-i\mu\right)\phi^\pm _{\bsk}(\tau)\Big|_{\tau=\tau_*}=\chi^\pm _{\bsk}\,,
\end{align}
where $\chi^\pm _{\bsk}$ are source terms introduced for each time path. Note that $\chi^\pm$ are independent of the boundary scalar fields $\bar{\phi}$ in the Dirichlet boundary problem. The conditions~\eqref{mixed_bc} kill the mode $\phi \sim (-\tau_*)^{\frac{d}{2}+i\mu}$ and pick up the mode $\phi \sim (-\tau_*)^{\frac{d}{2}-i\mu}$, which makes it possible to identify the resulting wavefunctions with CFT generating functions, as we now explain\footnote{One could pick up the other mode by imposing the boundary conditions,
\begin{align}
\left(\theta_\tau-\tfrac{d}{2}+i\mu\right)\phi^\pm_{\bsk}(\tau)\Big|_{\tau=\tau_*}=\chi^\pm_{\bsk}\,,
\end{align}
which can be realized by simply replacing $\mu\to-\mu$ in the following analysis. Also, one could impose different boundary conditions for each time path, but we employ the conditions~\eqref{mixed_bc} for technical simplicity.}.

\medskip
To make the variation problem consistent with the mixed boundary conditions~\eqref{mixed_bc}, let us modify the action~\eqref{dS_action_D} as
\begin{align}
S^\pm[\phi]&=S^\pm_{\rm bulk}[\phi]+S^\pm_{\rm bd}[\phi]\,,
\nonumber
\\
S^\pm_{\rm bulk}[\phi]&
=\frac{1}{2}
\int_{\mathcal{C}_\pm}
\frac{d\tau}{(-\tau)^{d+1}}\int \frac{d^d\bsk}{(2\pi)^d}\Big[
(\theta_\tau\phi_{-\bsk})(\theta_\tau\phi_{\bsk})
-(m^2+k^2\tau^2)\phi_{-\bsk}\phi_{\bsk}
\Big]\,,
\nonumber
\\
S^\pm_{\rm bd}[\phi]&=\frac{1}{2}\int \frac{d^d\bsk}{(2\pi)^d}(-\tau_*)^{-d}
\Big[
\left(\tfrac{d}{2}+ i\mu\right)
\phi_{-\bsk}\phi_{\bsk}
+2\phi_{-\bsk}\chi_{\bsk}^\pm
\Big]_{\tau=\tau_*}\,.\label{heavyaction1}
\end{align}
The classical solutions satisfying the mixed boundary conditions~\eqref{mixed_bc} and the Bunch-Davies conditions are given by
\begin{align}
\label{solution_mixed}
\phi^\pm_{\bsk}(\tau)=\mathcal{B}_M^\pm(k;\tau)\chi^\pm_{\bsk}
\end{align}
with the bulk-to-boundary propagators
\begin{align}
\mathcal{B}_M^+ (k;\tau)=\frac{(-\tau)^{\frac{d}{2}}H^{(2)}_{-i\mu}(-k\tau)}{(\theta_{\tau_*}-\frac{d}{2}-i\mu)\left[(-\tau_*)^{\frac{d}{2}}H^{(2)}_{-i\mu}(-k\tau_*)\right]}\,,
\\
\mathcal{B}_M^- (k;\tau)=\frac{(-\tau)^{\frac{d}{2}}H^{(1)}_{i\mu}(-k\tau)}{(\theta_{\tau_*}-\frac{d}{2}-i\mu)\left[(-\tau_*)^{\frac{d}{2}}H^{(1)}_{i\mu}(-k\tau_*)\right]}\,.
\end{align}
In the superhorizon regime $-k\tau_*\leq -k\tau\ll1$, they behave as
\begin{align}
\label{mixedBlimit+}
\mathcal{B}_M^+ (k;\tau_*)\simeq&
-(\tau/\tau_*)^{\frac{d}{2}-i\mu}\frac{1+e^{- \pi\mu}e^{i\alpha(\mu)}(-k\tau)^{2i\mu}}{2i\mu}\,,\\
\label{mixedBlimit-}
\mathcal{B}_M^- (k;\tau_*)\simeq&
-(\tau/\tau_*)^{\frac{d}{2}+i\mu}\frac{1+e^{- \pi\mu}e^{-i\alpha(\mu)}(-k\tau)^{-2i\mu}}{2i\mu \cdot e^{- \pi\mu}e^{-i\alpha(\mu)}(-k\tau_* )^{-2i\mu}}\notag\\
=&-(\tau/\tau_*)^{\frac{d}{2}-i\mu}\frac{1+e^{ \pi\mu}e^{i\alpha(\mu)}(-k\tau)^{2i\mu}}{2i\mu} \,.
\end{align}
One can see in Eq.~\eqref{mixedBlimit+} and the first line of Eq.~\eqref{mixedBlimit-} that the positive frequency modes in the denominator of Eq.~\eqref{dS_heavy_B} are projected out by the boundary conditions. Also, as the second expression of Eq.~\eqref{mixedBlimit-} shows, the two bulk-to-boundary propagators match with each other except for the Boltzmann factor $e^{\mp\pi\mu}$ in the numerator.

\medskip
Now the on-shell action with the solutions~\eqref{solution_mixed} reads
\begin{align}
S^{\pm}[\phi^\pm]&=\frac{1}{2}\left.\int \frac{d^d\bsk}{(2\pi)^d}(-\tau_*)^{-d}
\phi^\pm _{-\bsk}\chi^\pm _{\bsk}
\right|_{\tau=\tau_*}
\nonumber
\\
&=\frac{1}{2}\int \frac{d^d\bsk}{(2\pi)^d}
(-\tau_*)^{-d}
\mathcal{B}_M^\pm (k;\tau_*)
\chi_{\bsk}^\pm \chi^\pm _{-\bsk}
\nonumber
\\
&=\frac{1}{2}\int \frac{d^d\bsk}{(2\pi)^d}
(-\tau_*)^{-d}
\left[\frac{e^{\mp \pi\mu}e^{i\alpha(\mu)}(-k\tau_*)^{2i\mu}+1}{-2i\mu}
\right]
\chi^\pm_{\bsk}\chi^\pm_{-\bsk}
\,.\label{onshellmixed}
\end{align}
Then, the wavefunction $\Psi^\pm_M [\chi^\pm]$ with the mixed boundary conditions~\eqref{mixed_bc} reads
\begin{align}
\Psi^\pm_M [\chi^\pm]=e^{\pm iS^\pm [\phi^\pm]}=
\exp\left[
\mp \frac{1}{2}\int \frac{d^d\bsk}{(2\pi)^d}(-\tau_*)^{-d}\,
\frac{e^{
\mp \pi\mu}e^{i\alpha(\mu)}(-k\tau_*)^{2i\mu}+1}{2\mu}
\chi^\pm _{\bsk}\chi^\pm _{-\bsk}
\right]\,.\label{wavefmixedheavy}
\end{align}
Hence, the dual QFT two-point functions are given by
\begin{align}
\langle O_{\bsk}O_{-\bsk}\rangle'_{M\pm} =\mp (-\tau_*)^{-d}\frac{e^{
\mp \pi\mu}e^{i\alpha(\mu)}(-k\tau_*)^{2i\mu}+1}{2\mu}\,.\label{heavyCFT2ptM}
\end{align}
Note that the second term in the numerator is analytic, so that we may eliminate it by adding an extra boundary term $\sim \chi^\pm _{-\bsk}\chi^\pm _{\bsk}$ to the action.
More explicitly, we define the renormalized action as
\begin{align}
\widetilde{S}^\pm[\phi]&=S^\pm_{\rm bulk}[\phi]+\widetilde{S}^\pm_{\rm bd}[\phi]\,,
\nonumber
\\
\widetilde{S}^\pm_{\rm bd}[\phi]&=\frac{1}{2}\int \frac{d^d\bsk}{(2\pi)^d}(-\tau_*)^{-d}
\Big[
\left(\tfrac{d}{2}+ i\mu\right)
\phi_{-\bsk}\phi_{\bsk}
+2\phi_{-\bsk}\chi_{\bsk}^\pm
+(2i\mu)^{-1}\chi_{-\bsk}^\pm\chi_{\bsk}^\pm
\Big]_{\tau=\tau_*}\,,
\end{align}
where $S^\pm_{\rm bulk}[\phi]$ are given in the second line of \eqref{heavyaction1}. Note that the newly added counterterms do not affect the boundary conditions.
As a result, the renormalized wavefunctions $\widetilde{\Psi}^\pm_M [\chi^\pm]$ are given by
\begin{align}
\widetilde{\Psi}^\pm_M [\chi^\pm]=e^{\pm i\widetilde{S}^\pm [\phi^\pm]}=
\exp\left[
\mp \frac{1}{2}\int \frac{d^d\bsk}{(2\pi)^d}
\frac{e^{
\mp \pi\mu}e^{i\alpha(\mu)}}{2\mu}(-\tau_*)^{-d}(-k\tau_*)^{2i\mu}
\chi^\pm _{\bsk}\chi^\pm _{-\bsk}
\right]\,.\label{wavefmixedheavy2}
\end{align}
As advertised, the coefficients in the wavefunctions are naturally identified with conformal two-point functions of the dual operator $O$ with the conformal dimension $\Delta=\frac{d}{2}+i\mu$:
\begin{align}
\langle O_{\bsk}O_{-\bsk}\rangle'_{{\rm CFT}\pm} =\mp \frac{e^{\mp\pi\mu}e^{i\alpha(\mu)}}{2\mu}(-\tau_*)^{-d}(-k\tau_*)^{2i\mu}\,.\label{heavyCFT2pt}
\end{align}
Note that the dual CFTs are non-unitary since the conformal dimension $\Delta$ is complex.

\subsection{Double-trace deformations}\label{double_trace_def}

We have argued that the renormalized wavefunctions $\widetilde{\Psi}^\pm_M [\chi^\pm]$ with the mixed boundary conditions may be identified with CFT generating functions. In this subsection, we demonstrate that the wavefunctions $\Psi^\pm_D [\bar{\phi}]$ with the Dirichlet boundary conditions are identified with the generating functions of QFTs on a cyclic RG flow induced by double-trace deformations.

\medskip
To see this, let us first notice that $\Psi^\pm_D [\bar{\phi}]$ and $\widetilde{\Psi}^\pm_M [\chi^\pm]$ are related to each other by
\begin{align}
\label{Fourier}
\Psi^\pm _D [\bar{\phi}]=\!\int[d\chi^\pm ]\,\widetilde{\Psi}^\pm _M [\chi^\pm ]
\exp\left[\mp\frac{i}{2}\!\int\!\!\frac{d^d\bsk}{(2\pi)^d}
\frac{
\left(\tfrac{d}{2}+ i\mu\right)
\bar{\phi}_{-\bsk}\bar{\phi}_{\bsk}
+2\bar{\phi}_{-\bsk}\chi^\pm_{\bsk}
+(2i\mu)^{-1}\chi^\pm_{-\bsk}\chi^\pm_{\bsk}
}{(-\tau_*)^{d}}\right]\,.
\end{align}
According to the result of the last subsection, we are free to identify $\widetilde{\Psi}^\pm_M[\chi^\pm]$ with CFT generating functions $Z_{\rm CFT}^\pm[\chi^\pm]$ as 
\begin{align}
\widetilde{\Psi}^\pm_M[\chi^\pm]=Z_{\rm CFT}^\pm[\chi^\pm]=\int [d\Phi^\pm]\exp\left[-S^\pm_{\rm CFT}+\int \frac{d^d\bsk}{(2\pi)^d} \chi^\pm_{-\bsk}O_{\bsk}\right]\,,
\end{align}
where we schematically introduced the path integral for the would-be dual CFTs. In particular, $S^\pm_{\rm CFT}$ are CFT actions and $[d\Phi^\pm]$ is the path integral measure. The standard holographic dictionary says that $\chi^\pm$ source the dual CFT operator $O$. Substituting this into Eq.~\eqref{Fourier} gives a CFT interpretation of the wavefunctions $\Psi^\pm_D[\bar{\phi}]$:
\begin{align}
\Psi^\pm_D [\bar{\phi}]=\int [d\Phi^\pm]\, \exp\left[-S^\pm_{\rm QFT}\mp\frac{i}{2}\int\frac{d^d\bsk}{(2\pi)^d}
\left(\frac{
\tfrac{d}{2}- i\mu
}{(-\tau_*)^{d}}\,\bar{\phi}_{-\bsk}\bar{\phi}_{\bsk}\pm4\mu \bar{\phi}_{-\bsk}O_{\bsk}\right)\right]\,,\label{Fourier2}
\end{align}
where we defined
\begin{align}
S^\pm_{\rm QFT}=S^\pm_{\rm CFT}\mp\mu\left(-\tau_*\right)^d\int\frac{d^d\bsk}{(2\pi)^d}
O_{-\bsk}O_{\bsk}\,.
\end{align}
Note that now $\bar{\phi}$ plays the role of the source\footnote{
Note that in holographic renormalization, one may add arbitrary local counterterms of the source $\bar{\phi}$ to the boundary action. The $\bar{\phi}^2$ term in Eq.~\eqref{Fourier2} can be understood as such a counterterm.}.
Thus the correlation functions of the dual QFT can be written as 
\begin{align}
\langle\,\ldots\,\rangle_{D\pm}=\left\langle\,\ldots\,\exp\left(\pm\mu\left(-\tau_*\right)^d\int\frac{d^d\bsk}{(2\pi)^d}O_{-\bsk}O_{\bsk}\right)\right\rangle_{{\rm CFT}\pm}\,.
\end{align}
This shows that the QFTs dual to $\Psi^\pm_D$ are obtained by double-trace deformations of the CFTs dual to $\widetilde{\Psi}_M^\pm$.
Also the corresponding RG flow is cyclic because the double-trace operator $O^2$ has a conformal dimension $d+2i\mu$, which is also implied by the pure imaginary scaling of the two-point function \eqref{heavyCFT2pt}. This is analogous to the AdS case with a bulk scalar below the BF bound~\cite{Breitenlohner:1982jf}.

\subsection{Mixed boundary conditions for light fields}
\label{subsec:light}

For comparison, we discuss mixed boundary conditions for light scalars in the complementary series $0\leq m<\frac{d}{2}$, following Ref.~\cite{Das:2013qea}. In contrast to the heavy field case, the wavefunctions evaluated with mixed boundary conditions can be identified with CFT generating functions and the scalar field sources the shadow operator.
For technical simplicity, we focus on the mass range $\sqrt{\frac{d^2}{4}-1}< m<\frac{d}{2}$  in this section. More general cases are discussed in Appendix~\ref{app:light}. Readers familiar with this topic may safely jump to the next section after checking Eq.~\eqref{Fourier_light}, which will be used when studying dS correlators.

\subsubsection{Dirichlet boundary conditions}
First, let us briefly summarize the Dirichlet boundary conditions for light fields, for which the bulk-to-boundary propagators read
\begin{align}
\mathcal{B}_D^+ (k;\tau)=\frac{(-\tau)^{\frac{d}{2}}H^{(2)}_{\nu}(-k\tau)}{(-\tau_*)^{\frac{d}{2}}H^{(2)}_{\nu}(-k\tau_*)}\,,
\quad
\mathcal{B}_D^- (k;\tau)=\frac{(-\tau)^{\frac{d}{2}}H^{(1)}_{\nu}(-k\tau)}{(-\tau_*)^{\frac{d}{2}}H^{(1)}_{\nu}(-k\tau_*)}\,,
\end{align}
where we introduced $\nu=\sqrt{\frac{d^2}{4}-m^2}$. In the superhorizon regime $-k\tau_*\leq -k\tau\ll1$, they behave as
\begin{align}
\label{dS_light_B}
\mathcal{B}_D^\pm (k;\tau)&\simeq(\tau/\tau_*)^{\frac{d}{2}- \nu}
\, .
\end{align}
Note that the mode $\sim (-\tau)^{\frac{d}{2}- \nu}$ dominates over the other in the superhorizon limit, which is analogous to the AdS case.
Consequently, the bulk-to-boundary operators have a simple asymptotic behavior \eqref{dS_light_B} outside the horizon.
Then, the wavefunctions are given by
\begin{align}
\Psi^\pm _D [\bar{\phi}]=&\exp
\left[
\mp\frac{i}{2}\int \frac{d^d\bsk}{(2\pi)^d}
(-\tau_*)^{-d}\,
\left[\theta_\tau\mathcal{B}_D^\pm(k;\tau)\right]_{\tau=\tau_*}
\bar{\phi}_{\bsk}\bar{\phi}_{-\bsk}
\right]
\nonumber
\\
=&\exp
\bigg[
\mp\frac{i}{2}\int \frac{d^d\bsk}{(2\pi)^d}
(-\tau_*)^{-d}\,
\notag\\
&\times
\left(
\left(\tfrac{d}{2}-\nu\right)+\mathcal{O}\left(k^{2} \tau_{*}^{2}\right)+e^{\pm i \pi \nu} \cdot 2 \nu \cdot \frac{\Gamma(-\nu)}{2^{2 \nu} \Gamma(\nu)}\left(-k \tau_{*}\right)^{2 \nu}+\cdots
\right)
\bar{\phi}_{\bsk}\bar{\phi}_{-\bsk}
\bigg]\,,
\end{align}
where the first two terms are analytic in $ (-k \tau_* )^2$ and pure imaginary. The dots stand for non-analytic terms which are subleading in the limit $k\tau_* \ll 1$. When we compute the wavefunction squared $|\Psi_D |^2=\Psi_D^+\Psi_D^-$, these pure imaginary terms disappear due to the projection onto the imaginary part of the action. In addition, we can also eliminate these analytic terms by adding proper boundary terms to the action.

\medskip
More explicitly, if we focus on the mass regime $\sqrt{\frac{d^2}{4}-1}< m<\frac{d}{2}$, which corresponds to $0<\nu<1$, holographic renormalization can be performed as
\begin{align}
\label{dS/CFT_light_hr}
\Psi^\pm _D [\bar{\phi}]=&\exp
\left[
\frac{1}{2}\int \frac{d^d\bsk}{(2\pi)^d}
\left(
\langle O_{\bsk}O_{-\bsk}\rangle'_{{\rm CFT}_D\pm}
\bar{\phi}_{\bsk}\bar{\phi}_{-\bsk}
\mp i\,\frac{(\tfrac{d}{2}-\nu)\bar{\phi}_{\bsk}\bar{\phi}_{-\bsk}}{(-\tau_*)^{d}}
\right)
\right]\,,
\end{align}
where the second term is the counterterm. Also we introduced
\begin{align}
\label{ODirichlet}
\langle O_{\bsk}O_{-\bsk}\rangle' _{{\rm CFT}_D\pm} = \mp i\,\frac{2 \nu e^{\pm i \pi \nu} \Gamma(-\nu)}{2^{2 \nu} \Gamma(\nu)}(-\tau_*)^{-d+2\nu}k^{2 \nu}\,,
\end{align}
which can be identified with the conformal two-point function of a primary field with conformal dimension $\Delta= \frac{d}{2}+\nu$. Note that for $\nu>1$, there appear more counterterms, but the final result~\eqref{ODirichlet} still remains valid.

\subsubsection{Mixed boundary conditions}\label{mixedlight}

Next we study mixed boundary conditions:
\begin{align}
\label{mixed_bc_light}
\left(\theta_\tau-\frac{d}{2}+\nu \right)\phi^\pm_{\bsk}(\tau)\Big|_{\tau=\tau_*}=\chi^\pm_{\bsk}\,,
\end{align}
which kill the leading mode $\phi \sim (-\tau_* )^{\frac{d}{2}-\nu}$ and keep the mode $\phi \sim (-\tau_* )^{\frac{d}{2}+\nu}$. Together with the Bunch-Davies conditions, the classical solutions are given by
\begin{align}
\phi^\pm _{\boldsymbol{k}} (\tau)=\mathcal{B}_M^\pm (k ; \tau) \chi^\pm _{\boldsymbol{k}}
\end{align}
with the bulk-to-boundary propagators,
\begin{align}
\label{lightmixedB+1}
\mathcal{B}_M^+ (k;\tau)=&\frac{(-\tau)^{\frac{d}{2}}H^{(2)}_{\nu}(-k\tau)}{(\theta_{\tau_*}-\frac{d}{2}+\nu)\left[(-\tau_*)^{\frac{d}{2}}H^{(2)}_{\nu}(-k\tau_*)\right]}\,,\\
\mathcal{B}_M^- (k;\tau)=&\frac{(-\tau)^{\frac{d}{2}}H^{(1)}_{\nu}(-k\tau)}{(\theta_{\tau_*}-\frac{d}{2}+\nu)\left[(-\tau_*)^{\frac{d}{2}}H^{(1)}_{\nu}(-k\tau_*)\right]}\,.\label{lightmixedB-1}
\end{align}
In the superhorizon regime $-k\tau_*\leq -k\tau\ll1$, the bulk-to-boundary propagators behave as
\begin{align}
\mathcal{B}_M^\pm(k;\tau_*)\simeq
\frac{2^{2\nu}\Gamma(\nu)}{2\nu e^{\pm i\pi\nu}\Gamma(-\nu)}(-k\tau_*)^{-2\nu}\,,\label{lightBlimit}
\end{align}
where we assumed $0<\nu<1$. To make the variation problem consistent with the modified mixed boundary conditions~\eqref{mixed_bc_light}, we add a boundary term to the action~\eqref{dS_action_D} as follows:
\begin{align}
S^\pm_{\rm bd}[\phi]&=\frac{1}{2}\int \frac{d^d\bsk}{(2\pi)^d}(-\tau_*)^{-d}
\bigg[
\left(\tfrac{d}{2}-\nu \right)
\phi_{-\bsk}\phi_{\bsk}
+2\phi_{-\bsk}\chi^\pm_{\bsk}
\bigg]_{\tau=\tau_*}\,.\label{lightbdmixed}
\end{align}
Thus, the on-shell action reads
\begin{align}
S^\pm [\phi^\pm ]
=\frac{1}{2}\int \frac{d^d\bsk}{(2\pi)^d}
(-\tau_*)^{-d}
\left[\frac{2^{2 \nu} \Gamma(\nu)}{2\nu e^{\pm i \pi \nu} \Gamma(-\nu)}(-k\tau_* )^{-2\nu}+\cdots \right]
\chi^\pm_{\bsk}\chi^\pm_{-\bsk}
\,,
\end{align}
where the dots stand for subleading terms in the limit $k\tau_* \ll 1$. The corresponding wavefunctions $\Psi^\pm _M [\chi^\pm ]$ are
\begin{align}
\Psi^\pm _M [\chi^\pm ]=
\exp\left[
\pm \frac{i}{2}\int \frac{d^d\bsk}{(2\pi)^d}
\frac{2^{2 \nu} \Gamma(\nu)}{2\nu e^{\pm i \pi \nu} \Gamma(-\nu)}(-\tau_*)^{-d}(-k\tau_*)^{-2\nu}
\chi^\pm_{\bsk}\chi^\pm _{-\bsk}
\right]\,.\label{waveflightmixed}
\end{align}
The coefficient in the wavefunction can be identified with conformal two-point function as
\begin{align}
\langle O_\bsk O_{-\bsk}\rangle'_{{\rm CFT}_M\pm} =\pm i\,\frac{ 2^{2 \nu} \Gamma(\nu)}{2\nu e^{\pm i \pi \nu} \Gamma(-\nu)}(-\tau_*)^{-d-2\nu}k^{-2\nu}\,,
\end{align}
where the operator $O$ has a scaling dimension $\Delta=\frac{d}{2}-\nu$. This corresponds to the shadow operator of the CFT operator which appeared in Eq.~\eqref{ODirichlet} in the context of the Dirichlet boundary problem. Notice that in contrast to the heavy field case, no counterterm $\sim\chi^2$ in the boundary action is needed to have conformal two-point functions.

\subsubsection{Relation between the two CFTs}

Finally, we discuss how the two CFTs dual to the Dirichlet boundary condition and the mixed boundary condition are related with each other. For this purpose, let us first note the relation,
\begin{align}
\label{Fourier_light}
\Psi^\pm _D [\bar{\phi}]=\int [d\chi^\pm ]\,\Psi^\pm _M [\chi^\pm ]\exp\left[\mp\frac{i}{2}\int\frac{d^d\bsk}{(2\pi)^d}
\frac{
\left(\tfrac{d}{2}-\nu\right)
\bar{\phi}_{-\bsk}\bar{\phi}_{\bsk}
+2\bar{\phi}_{-\bsk}\chi^\pm_{\bsk}
}{(-\tau_*)^{d}}\right]\,,
\end{align}
which is analogous to Eq.~\eqref{Fourier} in the heavy field case. We also identify $\Psi^\pm_{D}[\bar{\phi}]$ and $\Psi^\pm_{M}[\bar{\phi}]$ with CFT generating functions $Z^\pm_{{\rm CFT}_D}[\bar{\phi}]$ and $Z^\pm_{{\rm CFT}_M}[\bar{\phi}]$ as
\begin{align}
\label{Psi_DZ_D}
\Psi^\pm _D [\bar{\phi}]&=Z^\pm_{{\rm CFT}_D}[\bar{\phi}]
\exp
\left[
\mp\frac{i}{2}\int \frac{d^d\bsk}{(2\pi)^d}
\frac{(\tfrac{d}{2}-\nu)\bar{\phi}_{\bsk}\bar{\phi}_{-\bsk}}{(-\tau_*)^{d}}
\right]
\,,
\\
\Psi^\pm _M [\chi^\pm]&=Z^\pm_{{\rm CFT}_M}[\chi^\pm]
\,,
\end{align}
where the second factor on the right hand side of the first equation reflects holographic renormalization (see discussion around Eq.~\eqref{dS/CFT_light_hr}). 
The subscripts $D$ and $M$ indicate that these CFTs are dual to the Dirichlet and the mixed boundary conditions, respectively. Then, the relation~\eqref{Fourier_light} implies that the two CFT generating functions are related with each other through the Legendre transformation\footnote{
Notice here that the $\bar{\phi}^2$ term in Eq.~\eqref{Fourier_light} and Eq.~\eqref{Psi_DZ_D} cancel out each other and then we arrive at a simple relation~\eqref{Z_DZ_M}. As we mentioned earlier, for $\nu>1$, there appear more counterterms in the Dirichlet boundary problem. Also, the mixed boundary condition~\eqref{mixed_bc_light} is modified as we discuss in Appendix~\ref{app:light}. Accordingly, Eqs.~\eqref{Fourier_light}-\eqref{Psi_DZ_D} acquire more counterterms. However, using the results in Appendix~\ref{app:light}, one can easily show that these counterterms cancel each other out and the relation~\eqref{Z_DZ_M} holds even for $\nu>1$.},
\begin{align}
\label{Z_DZ_M}
Z^\pm _{{\rm CFT}_D} [\bar{\phi}]=\int [d\chi^\pm ]\,Z^\pm _{{\rm CFT}_M}  [\chi^\pm ]\exp\left[\mp i\int\frac{d^d\bsk}{(2\pi)^d}
\frac{
\bar{\phi}_{-\bsk}\chi^\pm_{\bsk}
}{(-\tau_*)^{d}}\right]\,.
\end{align}
In particular, this directly shows that two-point functions in the two CFTs are inverse of each other,
\begin{align}
\langle O_\bsk O_{-\bsk}\rangle'_{{\rm CFT}_D\pm} =(-\tau_*)^{-2d}\Big(\langle O_\bsk O_{-\bsk}\rangle'_{{\rm CFT}_M\pm} \Big)^{-1}\,,
\end{align}
as we observed earlier. One can also show explicitly that the two CFTs are connected by an RG flow generated by double-trace deformations~\cite{Das:2013qea}, generalizing the AdS results in Ref.~\cite{Witten:2001ua,Berkooz:2002ug,Gubser:2002zh,Gubser:2002vv,Hartman:2006dy} via an appropriate analytic continuation.

\section{Correlation functions}
\label{Sec:correlators}

We have seen that wavefunctions of heavy fields with Dirichlet boundary conditions can be interpreted as the generating functions of the dual QFTs that are double-trace deformations of the CFTs dual to wavefunctions with the mixed boundary conditions. In this section we use the relation to express equal-time correlation functions on de Sitter space (which are normally computed with wavefunctions with Dirichlet boundary conditions) in terms of dual conformal correlators which appear in wavefunctions with the mixed boundary condition. This provides a dS/CFT dictionary for heavy fields in particular.

\subsection{General formula}

First, we derive a formula relating equal-time correlation functions on de Sitter space to
the wavefunctions with mixed boundary conditions. Recall that equal-time correlators of an operator $F[\phi_{\bsk}(\tau)]$ at $\tau=\tau_*$ are given by \cite{Maldacena:2002vr}
\begin{align}
\langle F[\phi_{\bsk}(\tau_*)]\rangle_{\rm dS}=\int [d\bar{\phi}]\left|\Psi_D [\bar{\phi}]\right|^2 F[\bar{\phi}_{\bsk}]
\quad
{\rm with}
\quad
\left|\Psi_D[\bar{\phi}]\right|^2=\Psi_D^+[\bar{\phi}]\Psi_D^-[\bar{\phi}]\,,
\label{Oexp}
\end{align}
where $F[\phi_{\bsk}(\tau)]$ is made from products of the field $\phi$ at the equal time $\tau$. For example, we have $F[\phi_{\bsk}(\tau)]=\phi_\bsk(\tau)\phi_{\bsk'}(\tau)$ for the computation of two-point functions $\langle\phi_{\bsk}(\tau_*)\phi_{\bsk'}(\tau_*)\rangle_{\rm dS}$.

\medskip
Next, we rewrite $\Psi_D^\pm$ in terms of the wavefunctions with mixed boundary conditions. For the heavy field case, we may consider either $\Psi_M^\pm$ or the renormalized version $\widetilde{\Psi}_M^\pm$ dual to the CFT generating functions. Here we consider the former for convenience. Similarly to Eq.~\eqref{Fourier}, ${\Psi}_D^\pm$ and ${\Psi}_M^\pm$ are related to each other as
\begin{align}
\Psi_D^\pm[\bar{\phi}]=\int [d\chi^\pm]\,\Psi_M^\pm[\chi^\pm]
\exp\left[\mp\frac{i}{2}\int\frac{d^d\bsk}{(2\pi)^d}
\frac{
\left(\tfrac{d}{2}+i\mu\right)
\bar{\phi}_{-\bsk}\bar{\phi}_{\bsk}
+2\bar{\phi}_{-\bsk}\chi_{\bsk}^\pm
}{(-\tau_*)^{d}}\right]\,.\label{wavefunctionphi}
\end{align}
The wavefunction squared then reads
\begin{align}
&\left|\Psi_D[\bar{\phi}]\right|^2=\int [d\chi^+ d\chi^-]\, \Psi^+ _M[\chi^+]\Psi^- _M[\chi^-]
\exp\left[
-i\int \frac{d^d\bsk}{(2\pi)^d}\frac{(\chi^+ _{-\bsk}-\chi^- _{-\bsk})\bar{\phi}_{\bsk}}{(-\tau_*)^{d}}
\right]\,,
\label{wavefunctionphi_squared}
\end{align}
where notice that the $\bar{\phi}^2$ terms cancel out when the $+$ and $-$ contributions are multiplied.
Also, essentially because of this type of cancellation, the same relation~\eqref{wavefunctionphi_squared} holds in the light field case too, as one may explicitly check using Eq.~\eqref{Fourier_light} and its $\nu>1$ generalization~\eqref{Fourier_light_nu>1}. An important point is that both $\chi^+$ and $\chi^-$ are coupled to $\bar{\phi}$, and so $\Psi_M^+$ and $\Psi_M^-$ nontrivially mix with each other upon integration over $\bar{\phi}$. This reflects the closed time path nature of the in-in formalism. For this reason, it is convenient to introduce
\begin{align}
\chi_{R,\bsk}=\frac{\chi^+ _{\bsk}+\chi^- _{\bsk}}{2}\,,
\quad
\chi_{A,\bsk}=\chi^+ _{\bsk}-\chi^- _{\bsk}\,.
\end{align}
In this language, we write
\begin{align}
\left|\Psi_D[\bar{\phi}]\right|^2=\int[d\chi_A]
\exp\left[
-i\int \frac{d^d\bsk}{(2\pi)^d}\frac{\chi_{A,-\bsk}\bar{\phi}_{\bsk}}{(-\tau_*)^d}
\right]
\int[d\chi_R]\,
\Psi_M^+[\chi_R+\tfrac{1}{2}\chi_A]\Psi_M^-[\chi_R-\tfrac{1}{2}\chi_A]\,.
\label{wavefsquared}
\end{align}
Substituting Eq.~\eqref{wavefsquared} into Eq.~\eqref{Oexp} and performing the integration over $\bar{\phi}$, we obtain
\begin{align}
&\langle F[\phi_{\bsk}(\tau_*)]\rangle_{{\rm dS}}
\nonumber
\\
&=
\int [d\chi_A
d\bar{\phi}]\,
\exp\left[
-i\int \frac{d^d\bsk}{(2\pi)^d}\frac{\chi_{A,-\bsk}\bar{\phi}_{\bsk}}{(-\tau_*)^d}
\right]
F\left[\bar{\phi}_{\bsk}\right]
\int [d\chi_R]\,\Psi_M^+\left[\chi_R+\tfrac{1}{2}\chi_A\right]\Psi_M^-\left[\chi_R-\tfrac{1}{2}\chi_A\right]
\nonumber
\\
&=
\int [d\chi_A]\,
\delta(\chi_A)
F\left[-i(-\tau_*)^d\partial_{\chi_{A,-\bsk}}\right]
\int [d\chi_R]\,\Psi_M^+\left[\chi_R+\tfrac{1}{2}\chi_A\right]\Psi_M^-\left[\chi_R-\tfrac{1}{2}\chi_A\right]\,,
\label{correlatorformula_pre}
\end{align}
where we performed integration by parts over $\chi_A$ at the second equality. In other words, we have derived the following master formula:
\begin{align}
&\langle F[\phi_{\bsk}(\tau_*)]\rangle_{{\rm dS}}
\nn\\
&\quad=
\left.F\left[-i(-\tau_*)^d\partial_{\chi_{A,-\bsk}}\right]
\int [d\chi_R]\,\Psi_M^+\left[\chi_R+\tfrac{1}{2}\chi_A\right]\Psi_M^-\left[\chi_R-\tfrac{1}{2}\chi_A\right]\right|_{\chi_A=0}\,,
\label{correlatorformula}
\end{align}
where notice again that the integration over $\chi_R$ nontrivially mixes $\Psi_M^+$ and $\Psi_M^-$ reflecting the closed time path nature.

\subsection{Two-point functions}

We now apply the general formula \eqref{correlatorformula} to two-point functions. Suppose that the wavefunctions with mixed boundary conditions are given by
\begin{align}
\label{PsiMheavy}
\Psi_M^\pm[\chi^\pm]&=
\exp\left[
\frac{1}{2}\int \frac{d^d\bsk}{(2\pi)^d}
\langle O_\bsk O_{-\bsk} \rangle '_{M\pm}
\chi_{\bsk}^\pm\chi_{-\bsk}^\pm
\right]\,.
\end{align}
Then, we find
\begin{align}
&\ln\Big[\Psi_M^+\left[\chi_R+\tfrac{1}{2}\chi_A\right]\Psi_M^-\left[\chi_R-\tfrac{1}{2}\chi_A\right]\Big]
\nn\\
&\quad=
\int \frac{d^d\bsk}{(2\pi)^d}
\bigg[\,
\frac{1}{2}\Big(\langle O_\bsk O_{-\bsk} \rangle '_{M+}+\langle O_\bsk O_{-\bsk} \rangle '_{M-}\Big)
\left(\chi_{R,\bsk}\,\chi_{R,-\bsk}+\frac{1}{4}\chi_{A,\bsk}\,\chi_{A,-\bsk}\right)
\nn\\
&\qquad\qquad\quad\quad~\,\,
+\frac{1}{4}\Big(\langle O_\bsk O_{-\bsk} \rangle '_{M+}-\langle O_\bsk O_{-\bsk} \rangle '_{M-}\Big)
\left(\chi_{R,\bsk}\,\chi_{A,-\bsk}+\chi_{A,\bsk}\,\chi_{R,-\bsk}\right)\bigg]\,,
\end{align}
which implies
\begin{align}
&\int [d\chi_R]\,\Psi_M^+\left[\chi_R+\tfrac{1}{2}\chi_A\right]\Psi_M^-\left[\chi_R-\tfrac{1}{2}\chi_A\right]
\nn\\
&\quad=
\exp\Bigg[
\frac{1}{2}\int \frac{d^d\bsk}{(2\pi)^d}
\frac{\langle O_\bsk O_{-\bsk} \rangle '_{M+}\langle O_\bsk O_{-\bsk} \rangle '_{M-}}{\langle O_\bsk O_{-\bsk} \rangle '_{M+}+\langle O_\bsk O_{-\bsk} \rangle '_{M-}}
\chi_{A,\bsk}\,\chi_{A,-\bsk}
\Bigg]\,.
\end{align}
Therefore, the de Sitter two-point function reads
\begin{align}
\label{2pt_dictionary_mixed}
\langle \phi_\bsk (\tau_* )\phi_{-\bsk} (\tau_* )\rangle '_{\rm dS}&=
-(-\tau_*)^{2d}\frac{\langle O_\bsk O_{-\bsk} \rangle '_{M+}\langle O_\bsk O_{-\bsk} \rangle '_{M-}}{\langle O_\bsk O_{-\bsk} \rangle '_{M+}+\langle O_\bsk O_{-\bsk} \rangle '_{M-}}\,.
\end{align}
This is the holographic dictionary when mixed boundary conditions are employed. Below we show that this dictionary reproduces the correct two-point functions as it should be.

\subsubsection{Heavy fields}\label{heavyfree1}

Let us begin by the heavy field case:
\begin{align}
\label{PsiMheavy}
\Psi_M^\pm[\chi^\pm]&=
\exp\left[
\frac{1}{2}\int \frac{d^d\bsk}{(2\pi)^d}
\langle O_\bsk O_{-\bsk} \rangle '_{M\pm}
\chi_{\bsk}^\pm\chi_{-\bsk}^\pm
\right]\,,
\\
\langle O_{\bsk}O_{-\bsk}\rangle'_{M\pm} &=\langle O_{\bsk}O_{-\bsk}\rangle'_{{\rm CFT}\pm}\mp\frac{(-\tau_*)^{-d}}{2\mu}
\,,
\end{align}
where the conformal two-point functions are given by
\begin{align}
\langle O_{\bsk}O_{-\bsk}\rangle'_{{\rm CFT}\pm} &=\mp \frac{e^{\mp\pi\mu}e^{i\alpha(\mu)}}{2\mu}(-\tau_*)^{-d}(-k\tau_*)^{2i\mu}\,.
\label{conformal_2pt}
\end{align}
Then, the de Sitter two-point function reads
\begin{align}
&\langle \phi_\bsk (\tau_* )\phi_{-\bsk} (\tau_* )\rangle '_{\rm dS}
\nn\\
&\quad=(-\tau_*)^{2d}\frac{\left(\frac{(-\tau_*)^{-d}}{2\mu}-\langle O_\bsk O_{-\bsk} \rangle '_{{\rm CFT}+}\right) \left(\frac{(-\tau_*)^{-d}}{2\mu}+\langle O_\bsk O_{-\bsk} \rangle '_{{\rm CFT}-}\right)}{\langle O_\bsk O_{-\bsk} \rangle '_{{\rm CFT}+} +\langle O_\bsk O_{-\bsk} \rangle '_{{\rm CFT}-}}
\,.
\end{align}
Notice here that the analytic terms in the numerator (which were subtracted by holographic renormalization) make the expression in terms of conformal two-point functions less simple compared to the original dictionary~\eqref{2pt_dictionary_mixed}. However, such terms are necessary to reproduce the correct two-point functions. Indeed, using the expression~\eqref{conformal_2pt}, we find\footnote{It is convenient to use $\langle O_{\bsk}O_{-\bsk}\rangle'_{{\rm CFT}+}=-e^{-2\pi\mu}\langle O_{\bsk}O_{-\bsk}\rangle'_{{\rm CFT}-}$ in the intermediate step.}
\begin{align}
\langle \phi_\bsk (\tau_* )\phi_{-\bsk} (\tau_* )\rangle '_{\rm dS}
&=\frac{(-\tau_*)^d}{2\mu}\frac{|1+e^{-\pi \mu}e^{i\alpha}(-k\tau_*)^{2i\mu}|^2}{1-e^{-2\pi\mu}}\,,
\label{light2ptresult}
\end{align}
which agrees with the result in canonical quantization.

\subsubsection{Light fields}\label{lightcorrelator}

Finally, let us consider the light field case:
\begin{align}
\Psi_M^\pm[\chi^\pm]&=
\exp\left[
\frac{1}{2}\int \frac{d^d\bsk}{(2\pi)^d}
\langle O_\bsk O_{-\bsk} \rangle '_{M\pm}
\chi_{\bsk}^\pm\chi_{-\bsk}^\pm
\right]\,,
\\
\langle O_\bsk O_{-\bsk} \rangle '_{M\pm}&=\langle O_\bsk O_{-\bsk} \rangle '_{{\rm CFT}_M\pm}=\pm i\,\frac{2^{2 \nu} \Gamma(\nu)}{2\nu e^{\pm i \pi \nu} \Gamma(-\nu)}(-\tau_*)^{-d-2\nu}k^{-2\nu}\,,
\label{waveflight}
\end{align}
where recall that no counterterm term $\sim\chi^2$ in the boundary action was required in the light field case.
Using the dictionary~\eqref{2pt_dictionary_mixed},
we obtain
\begin{align}
\langle \phi_\bsk (\tau_* )\phi_{-\bsk} (\tau_* )\rangle ' _{\rm dS}=&-(-\tau_*)^{2d}\frac{\langle O_\bsk O_{-\bsk} \rangle '_{{\rm CFT}_M+} \langle O_\bsk O_{-\bsk} \rangle '_{{\rm CFT}_M-}}{\langle O_\bsk O_{-\bsk} \rangle '_{{\rm CFT}_M+} +\langle O_\bsk O_{-\bsk} \rangle '_{{\rm CFT}_M-}}\notag\\
=& (-\tau_*)^{d}\frac{2^{2\nu}\Gamma (\nu)^2}{4\pi}(-k\tau_*)^{-2\nu}
\,, \label{light2pt}
\end{align}
which is consistent with the result of canonical quantization. Notice that de Sitter two-point functions have the same scaling as the dual CFT two point functions $\langle O_\bsk O_{-\bsk} \rangle ' _{{\rm CFT}_M\pm}$ which appears in the wavefunctions with mixed boundary conditions.

\section{Discussion}
\label{Sec:discussion}

In this paper we revisited holographic interpretation of de Sitter wavefunctions for free scalars. We demonstrated that especially for heavy scalars in the principal series, mixed boundary conditions provide natural identification of wavefunctions with generating functions of the would-be dual CFTs. In particular, wavefunctions with Dirichlet boundary conditions are dual to QFTs on an RG flow generated by double-trace deformations. Since CFT operators dual to heavy fields have complex conformal dimensions, the resultant RG flow is cyclic. This is in sharp contrast to the light field case, for which double-trace deformations interpolate two CFTs associated with Dirichlet and mixed boundary conditions. Besides, we provided a new dS/CFT dictionary of two-point functions that are applicable when mixed boundary conditions are employed.

\medskip
There are several interesting future directions along the line of the present work. The first thing to do is to generalize our analysis of wavefunctions with mixed boundary conditions to interacting theories and also to theories with spinning fields. Especially for heavy fields, mixed boundary conditions manifest conformal symmetry of dual QFT correlators, which could be useful when studying symmetry and analytic structures of cosmological correlators in the context of the cosmological collider program (see, e.g.,~\cite{Chen:2009zp,Baumann:2011nk,Noumi:2012vr,Arkani-Hamed:2015bza,Isono:2018rrb,Arkani-Hamed:2018kmz,Kim:2019wjo,Sleight:2019mgd,Sleight:2019hfp,Isono:2019wex,Baumann:2019oyu,Baumann:2020dch,Pajer:2020wnj,Sleight:2020obc,Goodhew:2020hob,Cespedes:2020xqq,Pajer:2020wxk}). It would also be interesting to study relations between dS Witten diagrams and AdS Witten diagrams based on our findings. More conceptually, our results imply that wavefunctions defined with Dirichlet boundary conditions are dual to QFTs on RG limit cycles whenever we take into account heavy fields required by UV completion of gravity in the bulk. It could be an obstruction to constructing realistic dS/CFT setups in string theory, at least as long as the standard holographic dictionary is employed. It would be interesting to reconsider why it is difficult to turn on stringy mass in higher spin dS/CFT from this perspective, which could be useful in understanding de Sitter space in string theory.

\section*{Acknowledgements}

We wish to express our gratitude to Yoshio Kikukawa for fruitful discussion, encouragement and careful reading of the manuscript. We are grateful to Gary Shiu and Toshiaki Takeuchi for valuable discussion and collaborations on related topics.
We thank Daniel Baumann, Heng-Yu Chen, Yu-tin Huang, Marc Gillioz, Guilherme L. Pimentel, Kostas Skenderis and Charlotte Sleight for useful discussions, in particular during the workshop ``Cosmology Meets CFT Correlators" held at National Taiwan University and the online workshop ``Cosmological Correlators." We also thank the Yukawa Institute for Theoretical Physics at Kyoto University for hospitality during the online workshop YITP-W-20-03 ``Strings and Fields 2020,'' where H.M.L. presented the results in this paper. H.I. is supported in part by the ``CUniverse'' research promotion project by Chulalongkorn University (grant reference CUAASC). T.N. is supported in part by JSPS KAKENHI Grant Numbers JP17H02894 and JP20H01902.

\appendix

\section{Mixed boundary condition for general $\nu$}

\label{app:light}

In Sec.~\ref{subsec:light}, we provided the relation between the wavefunctions with the mixed boundary condition and the dual conformal two-point functions by analyzing the asymptotic (superhorizon) behavior of the bulk-to-boundary propagators. Here we extend the discussion to general values of $\nu$. When $\nu >1$, the superhorizon limit of the bulk-to-boundary propagators \eqref{lightmixedB+1} and \eqref{lightmixedB-1} become rather complicated. To see that, we use the series expansion of the Hankel functions:
\begin{align}
H_{\nu}^{(2)}(-k \tau)&=i\,\frac{2^{\nu} \Gamma(\nu)}{\pi} \sum_{n=0}^{\infty}\left[a_{n}(-k \tau)^{2 n-\nu}+b_{n}(-k \tau)^{2 n+\nu}\right]\,,\\
H_{\nu}^{(1)}(-k \tau)&=-i\,\frac{2^{\nu} \Gamma(\nu)}{\pi} \sum_{n=0}^{\infty}\left[a_{n}(-k \tau)^{2 n-\nu}+\widetilde{b}_{n}(-k \tau)^{2 n+\nu}\right]\,,
\end{align}
where the coefficients $a_n$, $b_n$ and $\widetilde{b}_n$ are given by
\begin{align}
&a_{n}=\frac{(-1)^{n}}{2^{2 n} n !(1-\nu)_{n}}\,, \\
&b_{n}=\frac{(-1)^{n}}{2^{2 n} n !(1+\nu)_{n}} \frac{e^{i \nu \pi} \Gamma(-\nu)}{2^{2 \nu} \Gamma(\nu)}\,,\label{anbn} \quad
\quad \widetilde{b}_{n}=\frac{(-1)^{n}}{2^{2 n} n !(1+\nu)_{n}} \frac{e^{-i \nu \pi} \Gamma(-\nu)}{2^{2 \nu} \Gamma(\nu)}
\end{align}
with the shifted factorial $(x)_n=x(x+1)\cdots (x+n-1)$.
In terms of them, the propagators \eqref{lightmixedB+1} and \eqref{lightmixedB-1} can be written as
\begin{align}
\mathcal{B}_M^+(k ; \tau)&=(\tau/\tau_*)^{\frac{d}{2}}\frac{\sum_{n=0}^\infty\left[a_{n}(-k \tau)^{2 n-\nu}+b_{n}(-k \tau)^{2 n+\nu}\right]}{\sum_{n=1}^\infty2n a_{n}(-k \tau_*)^{2 n-\nu}+\sum_{n=0}^\infty(2n+2\nu )b_{n}(-k \tau_*)^{2 n+\nu}}\,, \label{Bexpansion+}\\
\mathcal{B}_M^-(k ; \tau)&=(\tau/\tau_*)^{\frac{d}{2}}\frac{\sum_{n=0}^\infty\left[a_{n}(-k \tau)^{2 n-\nu}+\widetilde{b}_{n}(-k \tau)^{2 n+\nu}\right]}{\sum_{n=1}^\infty2na_{n}(-k \tau_*)^{2 n-\nu}+\sum_{n=0}^\infty(2n+2\nu )\widetilde{b}_{n}(-k \tau_*)^{2 n+\nu}}\,. \label{Bexpansion-}
\end{align}
We can now see that when $\nu>1$, in each denominator, the terms $2na_{n}(-k \tau_*)^{2 n-\nu}$ with $1\leq n\leq\lfloor\nu\rfloor$\footnote{$\lfloor\nu\rfloor$ is the floor function, namely the integer $k$ satisfying $k\leq\lfloor\nu\rfloor<k+1$.} dominate over $2\nu b_0(-k\tau_*)^\nu$.  As a result, in the superhorizon regime $-k\tau_*\leq -k\tau\ll1$, the leading terms of \eqref{Bexpansion+} and \eqref{Bexpansion-} become proportional to $(-k\tau_*)^{-2}$, in sharp contrast with the $\nu<1$ case. Note that this issue does not appear in the AdS/CFT case as long as we focus on CFT operators above the unitarity bound $\Delta>\frac{d}{2}-1$, i.e., $\nu<1$.

\medskip
In order to regain the desired leading behavior $\sim(k\tau_*)^{-2\nu}$, we need to kill these unwanted terms, but this forces us to change the boundary conditions.
Concretely, we modify the boundary condition \eqref{mixed_bc_light} in the following manner:
\begin{align}
\bigg(\theta_\tau-\frac{d}{2}+\nu -\sum_{n=1}^{\lfloor \nu\rfloor}c_n^\pm (-k\tau)^{2n} \bigg)\phi^\pm_{\bsk}(\tau)\Big|_{\tau=\tau_*}=\chi^\pm_{\bsk}\,,\label{bdmodified}
\end{align}
where $c_n^\pm$ are real constants. We shall show shortly that $c_n^\pm$ can always be chosen to eliminate the unwanted terms in the denominators of \eqref{lightmixedB+1} and \eqref{lightmixedB-1}, so that the terms proportional to $ (-k\tau_*)^{\nu}$ become leading again. Under the modification, the bulk-to-boundary propagators become
\begin{align}
\label{lightmixedB+2}
\mathcal{B}_M^+ (k;\tau)=&\frac{(-\tau)^{\frac{d}{2}}H^{(2)}_{\nu}(-k\tau)}{(\theta_{\tau_*}-\frac{d}{2}+\nu-\sum_{n=1}^{\lfloor \nu\rfloor}c_n^+ (-k\tau_*)^{2n})\left[(-\tau_*)^{\frac{d}{2}}H^{(2)}_{\nu}(-k\tau_*)\right]}\,,\\
\mathcal{B}_M^- (k;\tau)=&\frac{(-\tau)^{\frac{d}{2}}H^{(1)}_{\nu}(-k\tau)}{(\theta_{\tau_*}-\frac{d}{2}+\nu-\sum_{n=1}^{\lfloor \nu\rfloor}c_n^- (-k\tau_*)^{2n})\left[(-\tau_*)^{\frac{d}{2}}H^{(1)}_{\nu}(-k\tau_*)\right]}\,.\label{lightmixedB-2}
\end{align}
The coefficients $c_n^\pm$ can be determined by employing their recurrence relations. 
Let us first look into $c_n^+$. 
The denominator of \eqref{lightmixedB+2} is expanded as
\begin{align}
i\,\frac{2^{\nu} \Gamma(\nu)}{\pi}(-\tau_*)^{\frac{d}{2}}&\Bigg[\sum_{n=1}^\infty 2 n a_{n}\left(-k \tau_{*}\right)^{2 n-\nu}-\sum_{n=0}^{\lfloor \nu\rfloor}\sum_{m=1}^\infty a_{n}c_m^+ \left(-k \tau_{*}\right)^{2 n+2m-\nu}\notag\\
&+\sum_{n=0}^\infty(2 n+2 \nu) b_{n}\left(-k \tau_{*}\right)^{2 n+\nu}-\sum_{n=0}^{\lfloor \nu\rfloor}\sum_{m=1}^\infty b_{n}c_m^+ \left(-k \tau_{*}\right)^{2 n+2m+\nu}\Bigg]\,.
\end{align}
To cancel all terms dominating over $2\nu b_0 \left(-k \tau_{*}\right)^{\nu}$, we impose the following recurrence relation: for $1 \leq n \leq \lfloor \nu\rfloor$,
\begin{align}
\label{cn+rec}
2na_n -a_0 c_n^+ -\sum_{i=1}^{n-1}a_{n-i} c_i^+=0\,,
\end{align}
where notice that it is independent of $b_n$.
Since $a_0=1$, this immediately yields $c_1^+ =2a_1$, with which we can solve the recurrence relation. The solution $c_n^+$ is then given b
\begin{align}
c_n^+ =&2na_n -\sum_{i=1}^{n-1}a_{n-i} c_i^+ \notag\\
=&2na_n  -\sum_{i_1=1}^{n-1}a_{n-i_1} \left(2i_1 a_{i_1} -\sum_{i_2=1}^{i_1-1}a_{i_1-i_2} c_{i_2}^+ \right)\notag\\
\vdots&\notag\\
=&\sum_{m=1}^{n}\sum_{\substack{i_1 +i_2 \cdots+i_m =n\\i_1,i_2 ,\cdots, i_m \geq 1}}(-1)^{m+1}2 i_1 A_{i_1 i_2 \cdots i_m}\,,
\end{align}
where we introduced
\begin{align}
A_{i_1 i_2 \cdots i_m}\equiv a_{i_1}a_{i_2}\cdots a_{i_m}\,.
\end{align}
The coefficients $c_n^-$ can be fixed exactly in the same manner:
Since the unwanted terms in \eqref{Bexpansion-} are exactly the same as in \eqref{Bexpansion+}, the recurrence relation for $c_n^-$ takes exactly the same form as \eqref{cn+rec}. Therefore, we find $c_n^-=c_n^+$, which we denote by $c_n$ for simplicity. 

\medskip
Now, even for $\nu>1$, in the superhorizon regime, the bulk-to-boundary propagators behave again as
\begin{align}
\mathcal{B}_M^\pm (k;\tau_*)\simeq
\frac{2^{2\nu}\Gamma(\nu)}{2\nu e^{\pm i\pi\nu}\Gamma(-\nu)}(-k\tau_*)^{-2\nu}\,.
\end{align}
To make the variation problem consistent with the modified boundary conditions~\eqref{bdmodified}, the boundary action must be chosen to be
\begin{align}
S^\pm_{\rm bd}[\phi]&=\frac{1}{2}\int \frac{d^d\bsk}{(2\pi)^d}(-\tau)^{-d}
\bigg[
\bigg(\tfrac{d}{2}-\nu+\sum_{n=1}^{\lfloor \nu\rfloor}c_n (-k\tau)^{2n}\bigg)
\phi_{-\bsk}\phi_{\bsk}
+2\phi_{-\bsk}\chi^\pm_{\bsk}
\bigg]_{\tau=\tau_*}\,.\label{lightbdmixed2}
\end{align}
Namely, the modification of the boundary condition forces the extra local counterterms. As a result, we reproduce the wavefunction \eqref{waveflightmixed}, so that the results we derived in Sec \ref{mixedlight} apply to general value of $\nu$. On the other hand, the modification of the boundary actions \eqref{lightbdmixed2} affects the relation between the wavefunctions $\Psi_D^\pm$ and $\Psi_M^\pm$,
\begin{align}
\label{Fourier_light_nu>1}
\Psi^\pm _D [\bar{\phi}]&=\int [d\chi^\pm ]\,\Psi^\pm _M [\chi^\pm ] \nn\\
&\quad\times
\exp\Bigg[\mp\frac{i}{2}\int\frac{d^d\bsk}{(2\pi)^d}
\frac{
\left(\tfrac{d}{2}-\nu+\sum_{n=1}^{\lfloor \nu\rfloor}c_n^\pm (-k\tau_*)^{2n}\right)
\bar{\phi}_{-\bsk}\bar{\phi}_{\bsk}
+2\bar{\phi}_{-\bsk}\chi^\pm_{\bsk}
}{(-\tau_*)^{d}}\Bigg]\,.
\end{align}
The summation over $n$ in the exponent, however, cancels out in the wavefunction squared, thereby yielding the same form \eqref{wavefunctionphi_squared}.

\bibliography{reference}{}
\bibliographystyle{utphys}

\end{document}